\documentclass{article}

\usepackage{amssymb}
\usepackage{verbatim}
\usepackage[francais,english]{babel}
\usepackage{graphicx}
\usepackage{graphics}
\usepackage{lineno}
\usepackage[latin1]{inputenc}
\usepackage[table]{xcolor}
\usepackage[]{natbib}
\usepackage[]{multirow}
\usepackage[]{a4wide}
\usepackage{fancyhdr}
\usepackage{lastpage}
\begin{document}
%\pagenumbering{arabic}
%\cfoot{\thepage/\pageref{LastPage}}
%\setcounter{page}{1}

\baselineskip 0.8cm

\title{Arbitrariness of peer review: A Bayesian analysis of the NIPS experiment} 

\date{}

\author{Olivier Fran\c cois}

\maketitle

\begin{center}
Universit\'e Grenoble-Alpes, Centre National de la Recherche Scientifique, TIMC-IMAG UMR 5525, Grenoble, 38042, France.\\

\end{center}

\noindent{Running Title: Arbitrariness of peer review\\}


\vspace{1cm}

{\noindent Corresponding Author: Olivier Fran\c cois \\
Universit\'e Grenoble-Alpes,\\ 
TIMC-IMAG,  UMR CNRS 5525,\\
Grenoble, 38042, France.\\
+334 56 52 00 25 (Phone)  \\
+334 56 52 00 55 (Fax)  \\
 \texttt{olivier.francois@imag.fr}}

\clearpage
\newpage 

\begin{center}
{\bf Abstract}
\end{center}

The principle of peer review is central to the evaluation of research, by ensuring that only high-quality items are funded or published. But peer review has also received criticism, as the selection of reviewers may introduce biases in the system. In 2014, the organizers of the ``Neural Information Processing Systems\rq\rq{} conference conducted an experiment in which $10\%$ of submitted manuscripts (166 items) went through the review process twice. Arbitrariness was measured as the conditional probability for an accepted submission to get rejected if examined by the second committee. This number was equal to $60\%$, for a total acceptance rate equal to $22.5\%$. Here we present a Bayesian analysis of those two numbers, by introducing a hidden parameter which measures the probability that a submission meets basic quality criteria.  The standard quality criteria usually include novelty, clarity, reproducibility, correctness and no form of misconduct, and are met by a large proportions of submitted items. The Bayesian estimate for the hidden parameter was equal to $56\%$ ($95\%$CI: $ I = (0.34, 0.83)$), and had a clear interpretation. The result suggested the total acceptance rate should be increased in order to decrease arbitrariness estimates in future review processes.

\clearpage
\newpage 

\section{Introduction}

The principle of peer review is central to the evaluation of research proposals  and research studies, by ensuring that only high-quality items are funded or published. Since its origin, the aim of peer-review has been to filter out the lack of novelty, flaws in research methodology or data, lack of reproducibility, falsification, plagiarism and other forms of misconduct (Hames 2007; Vintzileos et al. 2010). But peer review has also been criticized on the grounds that it imposes burden on research communities, that the selection of reviewers may introduce biases in the system, and that the reviewers\rq{} judgements may be subjective or arbitrary (Kassirer and Campion 1994; Hojat et al. 2003; Li and Agha 2015).    Arbitrariness of peer review, which is the quality of accepting submitted items by chance or whim, and not by necessity or rationality, can be measured by the heterogeneity of evaluations among raters during the review process (Mutz et al. 2012; Marsh et al. 2008; Giraudeau et al. 2011).

In 2014, the organizers of the {\it Neural Information Processing Systems} (NIPS) conference, Corinna Cortes and Neil Lawrence, decided to look at how fair the conference evaluation system was (Langford and Guzdial 2015). NIPS is one of the main theoretical computer science conferences, and its review process has an advanced format which includes double blind review and the possibility of rebuttal for authors. Cortes and Lawrence ran the {\it NIPS experiment} in which $1/10$ of manuscripts (items) submitted to NIPS went through the review process twice. A total of $n = 166$ submissions were reviewed by two independent program committees, and the discrepancy of committee decisions was reported in a fully transparent way (Langford and Guzdial 2015).

The NIPS organizers defined {\it arbitrariness} as the conditional probability, $a$, for an accepted submission to get rejected if examined by a second committee. From the NIPS experiment, the observed arbitrariness was equal to $\hat a = 60\%$. Since the total acceptance rate was equal to $\hat \pi = 22.5\%$, the $60\%$ estimate was close to the maximal value of arbitrariness, $a = 77.5\%$.  So, what do these numbers mean? Is the NIPS review process unfair? Which practical guidelines could be deduced from these results for the organization of future conferences?  In this short survey, we propose to interpret the values of observed arbitrariness and acceptance rate using Bayesian data analysis (Gelman et al. 2014). We introduced a family of statistical models for the NIPS experiment data. These models include a hidden parameter, $x$, corresponding to the probability that a submitted item meets the basic quality criteria. They assume that items which do not meet those minimal quality criteria are almost surely rejected by both committees. We used the models to quantify uncertainty on the observed value of arbitrariness and on the hidden variable $x$, and computed model probabilities for several conditional acceptance rules.

\section{Explaining the arbitrariness measure with models}

Let $n$ be the total number of submissions during the NIPS experiment ($n = 166$), and let $\hat \pi = k/n = 22.5\%$ be the total acceptance rate at the NIPS conference. To explain an observed arbitrariness level, $a$, we introduce the {\it Reject or Flip a Coin (RFC)} model, which is based on two parameters, $x$ and $y$. The first parameter, $x$, is a hidden variable  representing the probability that an item meets basic quality criteria, such as novelty, clarity, absence of methodological flaws, reproducibility of results, and no form of misconduct. The second parameter, $y$, represents the conditional probability that an item meeting all quality criteria is accepted. Items that fail to meet all quality criteria are rejected with probability one.

From basic probability theory,  the total acceptance rate in the RFC model is equal to $\pi = xy$,  and arbitrariness is equal to $a = 1- y$. Thus,  model parameters are related through  the following relationship

$$
a = 1 - \frac{\pi}{x} \leq 1 - \pi \, .
$$  
When the total acceptance rate is known,  the arbitrariness level is maximal for $x = 100\%$, and the maximum value is $1 - \pi$. Arbitrariness is avoided when $a = 0$. In this case, the total acceptance rate corresponds to the rate of items meeting the quality criteria, and we have $y = 1$. With the NIPS experiment data,  the moment estimates of $x$ and $y$ are equal to $\hat x = 56\%$ and $\hat y = 40\%$ respectively. 

Assuming non-informative prior distributions for $x$ and $y$, we used the Bayes formula to derive the posterior distribution of the model parameter $(x,y)$. The posterior distribution can be described by the following equation
$$
p( x, y | \hat a,k,n) \propto (xy)^k (1 - xy)^{n-k}   y^{(1-\hat a)k}  (1 - y)^{\hat a k} \, .
$$
The moment estimates  $\hat x = 56\%$ and $\hat y = 40\%$ correspond to the mode of the posterior distribution and to the maximum of the likelihood function.  By integrating with respect to the variable $x$, we obtained the posterior distribution of $y$ as follows
$$
p( y |\hat a,k,n)  \propto   y^{(1-\hat a)k}  (1 - y)^{\hat a k} \int_0^1 (xy)^k (1 - xy)^{n-k} dx \propto y^{(1-\hat a)k - 1}  (1 - y)^{\hat a k} \, .
$$
In other words, $p(y | \hat a, k, n)$ is a beta distribution with parameters $(1-\hat a)k$ and $(\hat a k + 1)$
$$
 y | \hat a, k, n \sim {\rm beta}((1-\hat a)k , \hat a k + 1 ).
$$
To provide an exact simulation algorithm for the posterior distribution of the model parameter $(x,y)$, we computed the
 the density of the conditional distribution of $x$ given $y, \hat a, k,n$. This conditional distribution could be represented as the distribution of the random variable $x^\star/y$ where $x^\star$ is drawn from a beta$(k+1, n-k+1)$ distribution conditioned on being lower than $y$. 
 
Next, we used a basic rejection algorithm for sampling 100,000 replicates from the posterior distribution (Figure 1). A Bayesian Monte Carlo estimate of the arbitrariness parameter, $a = 1-y$, was $61\%$, and its 95$\%$ credibility interval was equal to $I =(0.43, 0.73)$. The Bayesian estimate for the rate of items, $x$, meeting all quality criteria was $56\%$, and the 95$\%$ credibility interval was $I = (0.34, 0.83)$ (Figure 2). 

\begin{figure}[ht]
\begin{center}
\includegraphics[width = 12cm]{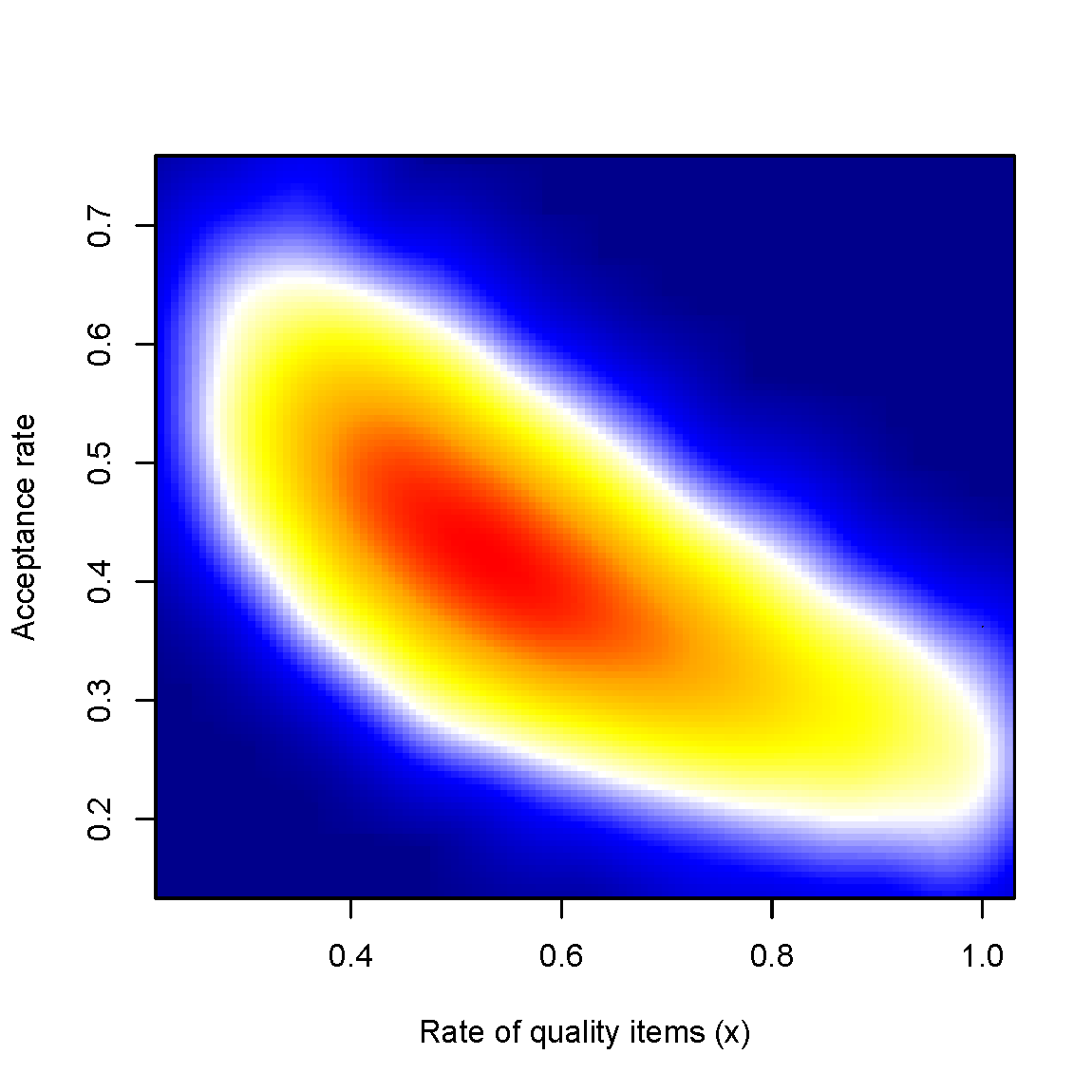}\\
\caption{Joint posterior distribution of the $x$ and $y$ parameters in the RFC model.}
\end{center}
\end{figure}

\begin{figure}[ht]
\begin{center}
\includegraphics[width = 12cm]{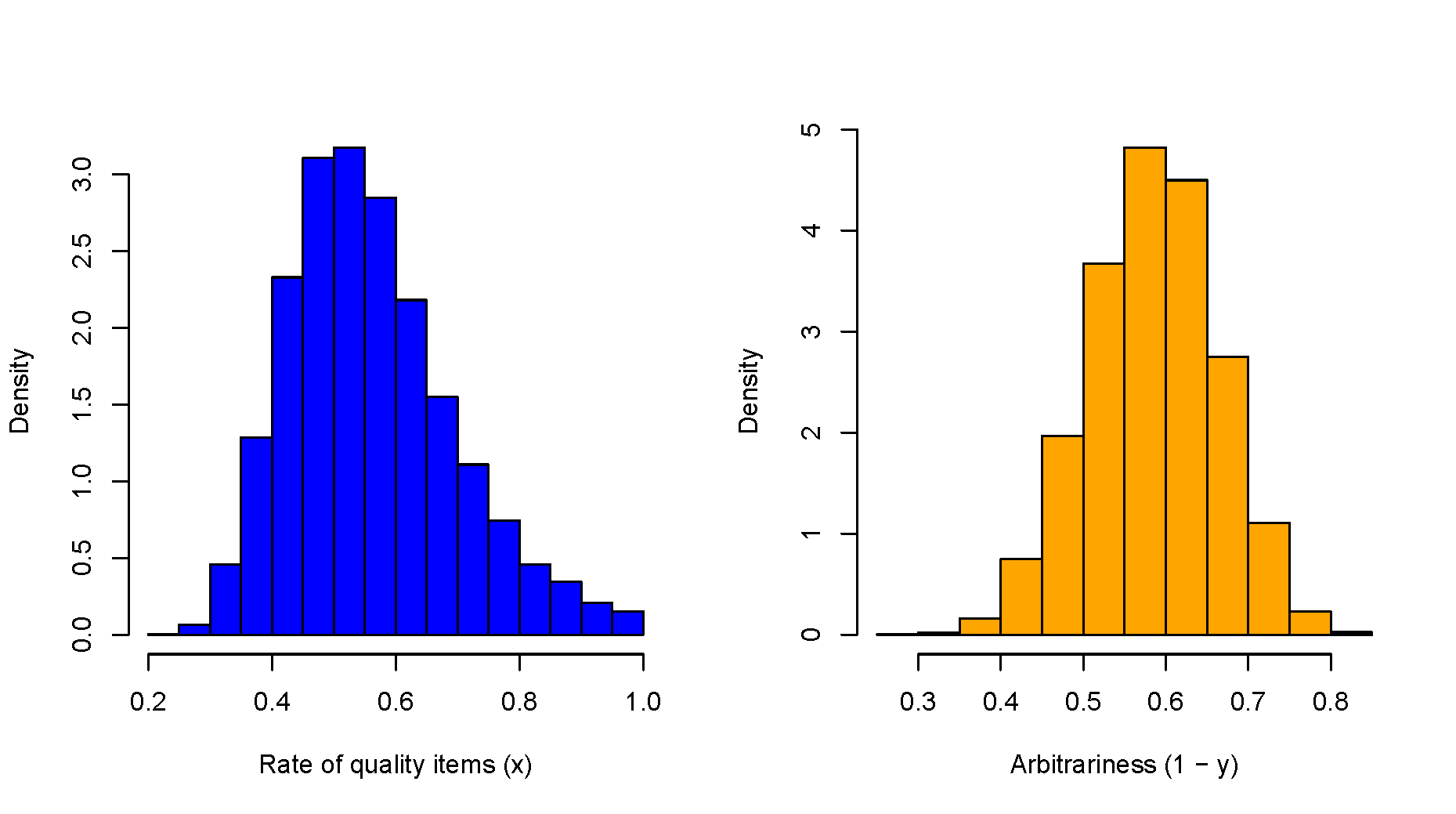}\\
\caption{Posterior density for the rate of items meeting all quality criteria, $x$, and for the arbitrariness parameter, $a = 1-y$, in the RFC model.}
\end{center}
\end{figure}

To model the fact that some submissions are clearly accepted by any committee, we considered an extension of the RFC model in which both low-quality and high-quality items give rise to deterministic decisions. The new model is called the {\it Reject, Accept or Flip a Coin (RAFC)} model. Low-quality items are rejected by committees with probability 1, whereas high-quality items are accepted with probability 1. The rate of low-quality items is $1 - x$, and the rate of high quality items is $\alpha x$, where $\alpha$ is a known parameter with value between 0 and 1. Items that meet all quality criteria but that are not high-quality items are accepted with probability $y$.  In the RAFC model, arbitrariness is parameterized as 
$$
a =  \frac{(1-\alpha)xy(1-y)}{ \alpha x + (1-\alpha)xy } \, ,
$$
and the total acceptance rate $\pi$ is equal to
$$
\pi =  \alpha x + (1-\alpha)xy  \, .
$$
The RFC model is a particular instance of the RAFC model obtained for $\alpha = 0$. The question here was to evaluate for which values of $\alpha$ the RAFC model could provide a better fit to the data than the RFC model. 

Assuming non-informative prior distributions for $x$ and $y$  in the RAFC model,  the posterior distribution of $(x,y)$ was described by the following formula
$$
\begin{array}{ll}
p( x, y |\hat a,k,n)   & \propto  (\alpha x + (1-\alpha)xy )^k  (1 - \alpha x - (1-\alpha)xy)^{n-k}  \\ 
& \times \left( (1-\alpha)xy(1-y) / ( \alpha x + (1-\alpha)xy )  \right)^{\hat a k} \\
& \times \left( 1 -  (1-\alpha)xy(1-y) / ( \alpha x + (1-\alpha)xy ) \right)^{(1 - \hat a ) k} 
 \end{array}
$$
Taking $\alpha = 5\%$, the mode of the posterior distribution corresponded to the parameter values $\hat x = 69\%$ and $\hat y = 29\%$. To sample from the posterior distribution and evaluate uncertainty on model parameters, we used an approximate Bayesian computation (ABC) approach based on 100,000 simulations (Csilléry et al. 2010). Using ABC and $\alpha = 5\%$, the Bayesian estimates of $x$ and $y$ were calculated as $\hat x = 67\%$ ($I = (.38, 1.0) $) and $\hat y = 32\%$ ($I = (.14, .54) $)  respectively. A Bayesian estimate of arbitrariness was equal to $57\%$, and its 95$\%$ credibility interval was equal to $I = (0.41, 0.63)$. 

In addition to $\alpha = 5\%$, five other values of $\alpha$ were tested ($\alpha=$ 0$\%$, 2.5$\%$, 10$\%$,20$\%$,50$\%$) and the corresponding RAFC models were compared with the model using $\alpha = 5\%$. The comparison was achieved by using an ABC approach to evaluate posterior model probabilities. ABC model choice indicated that smaller values of $\alpha$ provided better fit to the data than larger values. The RAFC model using $\alpha = 5\%$ corresponded to the highest posterior probability ($p=26\%$).  In a pairwise comparison with the RFC model, the RAFC model using $\alpha = 5\%$ had a probability of $56\%$, and the Bayes factor was equal to BF = 1.31 (barely worth mentioning).

\section{Discussion}

Peer review is not perfect, and levels of arbitrariness in the range $(0.43, 0.73)$ supported the evidence for biases during the review process. In light of the RFC model interpretation, the results indicated that the burdens on reviewers, which is one of the biggest costs in the peer review system, could be alleviated by restricting their role to check whether basic quality criteria such as novelty, absence of methodological flaws, reproducibility of results, are met. This phase of the review process should end with an acceptance rate, $x$, within the interval $(0.34, 0.83)$. In a second phase, flipping biased coins with success probability equal to $\pi/x$ would lead to the same acceptance rate and level of arbitrariness as in the NIPS experiment. It seems however unlikely that this apparently random process could be envisaged as an alternative to the original review process in future experiments.

One of the highest costs in the peer review system is for the submitters themselves and for their funding agencies. Arbitrary decisions delay publication or funding of research works that would deserve merit. Those decisions can have a negative influence on junior researchers who might be more importantly impacted by arbitrary rejection than senior researchers (see Bourne 2005, ``Rule 5: learn to live with rejection\rq\rq{}).  A positive aspect of the NIPS experiment  is that its analysis provides a way to restrict arbitrariness in future instances of the peer review process. The estimate of $\hat x = 56\%$ is a clear suggestion to push the total acceptance rate close to $\pi = 56\%$, so that arbitrariness would be closer to zero.

Many critics claim that review processes are unnecessary and slow the communication of information. Initiatives such as preprint repositories have demonstrated the utility of open science (Sitek and Bertelmann 2014). Some multidisciplinary open access journals use publication criteria based on ethical standards and the rigor of the methodology and conclusions reported. Although surveys of peer review among fee-charging open access journals showed that the target  of publishing \lq\lq{}scientifically rigorous research\rq{}\rq{} could be difficult to reach (Bohannon 2013), the lesson from the NIPS experiment is that accepting all scientifically rigorous research works would reduce arbitrariness to very small levels.

\section{References}

{~}

Bohannon J (2013). Who\rq{}s afraid of peer review? Science 342 (6154): 60-65.

Bourne PE (2005). Ten simple rules for getting published. PLoS Computational Biology 1(5): e57.

Csilléry K, Blum MGB, Gaggiotti OE, François O (2010). Approximate Bayesian computation (ABC) in practice. Trends in Ecology and Evolution 25(7):410-418.

Gelman A, Carlin JB, Stern HS,  Rubin DB (2014). {\it Bayesian Data Analysis}. London: Chapman Hall/CRC.

Giraudeau B, Leyrat C, Le Gouge A, Léger J, Caille A (2011). Peer review of grant applications: a simple method to identify proposals with discordant reviews. PLoS ONE 6: e27557.

Hames I (2007). {\it Peer Review and Manuscript Management in Scientific Journals: Guidelines for Good Practice}. Oxford, United Kingdom: Wiley-Blackwell.

Hojat M, Gonnella JS,  Caelleigh AS (2003). Impartial judgment by the gatekeepers of science: fallibility and accountability in the peer review process. Advances in Health Sciences Education 8(1):75-96.

Kassirer JP, Campion EW (1994). Peer review: crude and understudied, but indispensable. JAMA 272(2): 96-97.

Li D, Agha L (2015). Big names or big ideas: Do peer review panels select the best science proposals? Science 348:434-438.

Langford J, Guzdial M (2015) The arbitrariness of reviews, and advice for school administrators. Communications of the ACM 58 (4):12-13.

Marsh HW, Jayasinghe UW, Bond NW (2008). Improving the peer-review process for grant applications - reliability, validity, bias, and generalizability. American Psychologist 63: 160-168.

Mutz R, Bornmann L, Daniel H-D (2012). Heterogeneity of inter-rater reliabilities of grant peer reviews and its determinants: A general estimating equations approach. PLoS ONE. 7: e48509.

Sitek D, Bertelmann R (2014). Open access: A state of the art. In Sönke Bartling; Sascha Friesike. Opening Science. Springer. p. 139.

Vintzileos AM, Ananth, CV (2010). The art of peer-reviewing  an original research paper: Important tips and guidelines.
J Ultrasound Med (29):513-518.

\end{document}